\documentclass[
aip,
rsi,
amsmath,
amssymb,
preprint
]{revtex4-1}

\usepackage[
a4paper,
left=2cm,
right=5.0cm,
top=2.5cm,
bottom=2.5cm
]{geometry}

\usepackage[utf8]{inputenc}
\usepackage[T1]{fontenc}
\usepackage{graphicx}
\usepackage{dcolumn}
\usepackage{bm}
\usepackage{amsmath}
\usepackage{amssymb}
\usepackage{mathptmx}

\begin{document}
\title{Pulsed ion deflection to overcome detector saturation in cryogenic ice sampling}

\author{P. Samarth}
\email{samarth@mail.strw.leidenuniv.nl}
\affiliation{
Laboratory for Astrophysics, Leiden Observatory, Leiden University,
P.O. Box 9513, NL-2300 RA Leiden, The Netherlands
}

\author{M. Bulak}
\affiliation{
Laboratory for Astrophysics, Leiden Observatory, Leiden University,
P.O. Box 9513, NL-2300 RA Leiden, The Netherlands
}

\author{D. Paardekooper}
\affiliation{
Laboratory for Astrophysics, Leiden Observatory, Leiden University,
P.O. Box 9513, NL-2300 RA Leiden, The Netherlands
}

\author{K.-J. Chuang}
\affiliation{
Laboratory for Astrophysics, Leiden Observatory, Leiden University,
P.O. Box 9513, NL-2300 RA Leiden, The Netherlands
}

\author{H. Linnartz}
\affiliation{
Laboratory for Astrophysics, Leiden Observatory, Leiden University,
P.O. Box 9513, NL-2300 RA Leiden, The Netherlands
}

\date{Received 6 November 2023; accepted 12 January 2024}

\begin{abstract}
In 2014, we introduced a new experimental approach to study the UV
photo-processing of cryogenic ices of astrophysical interest using laser
ablation in combination with ionization and time-of-flight mass spectrometry
(ToF-MS). The setup, MATRI$^2$CES (Mass Analytical Tool to Research
Interstellar ICES), allowed us to detect newly formed species at low
abundances. However, we found that with increasing molecular complexity,
the detection of larger photoproducts was hindered by the dynamic range of
the detector. Here, we introduce a method to overcome this issue that we
expect to be useful for similar applications in other research fields.

The concept is based on a precisely controlled high-energy pulser that
regulates the voltage across the deflection plates of the ToF-MS to deflect
the most abundant species and prevent them from reaching the detector.
In this way, the detector sensitivity can be increased from an operating
voltage of 2500~V up to 3000~V. The applicability is first illustrated
using an argon matrix, in which $^{40}$Ar$^+$ ions are deflected to increase
the detection sensitivity for $^{40}$Ar$^{2+}$ at $m/z=20$ and
$^{40}$Ar$_2^+$ at $m/z=80$ by a factor of 30. Similarly, we show that
substantially larger complex organic molecules can be measured in
UV-irradiated methanol ice.
\end{abstract}

\maketitle

\section{\label{sec:level1} Introduction\protect}

 As of Winter 2023, around 300 distinct molecular species, including dozens of so-called complex organic molecules (COMs), have been identified in the interstellar and circumstellar medium. The majority of these species have been identified as gas-phase species in the microwave or sub-millimeter region using ground-based telescopes, such as ALMA, IRAM30m, GBT, or YEBES40m \cite{McGuire_2022, belloche_re-exploring_2019, guelin_organic_2022}. It is generally accepted that a substantial amount of these COMs is formed on icy dust grains. Indeed, a dozen of different ice constituents has been observationally confirmed, comprising mainly smaller and abundant species that upon (non)energetic processing act as precursors in chemical pathways to molecular complexity \cite{Yang_Green_Pontoppidan_Bergner_Cleeves_II_Garrod_Jin_Kim_Kim_etal_2022, McClure_Rocha_Pontoppidan_Crouzet_Chu_Dartois_Lamberts_Noble_Pendleton_Perotti_etal_2023}. Over the last few decades, laboratory efforts have been made to study such solid-state processes under fully controlled conditions that are similar to the interstellar environment. 
 The chemistry representative for interstellar ice mantles is typically investigated in ultra-high vacuum setups, operating with base pressures in the 10$^{-9}$ to 10$^{-11}$ mbar regime. Ices are grown with thicknesses ranging from sub-monolayer to several tens of monolayers on top of cryo-cooled and non-reactive substrates, typically in the 10-20 K range. These ices are then exposed to different forms of processing, such as impacting atoms, electrons, photons, and cosmic rays or irradiation by UV light, that simulates the interstellar radiation field experienced by these mantles \cite{Chuang_Fedoseev_Qasim_Ioppolo_Jger_Henning_Palumbo_Van_Dishoeck_Linnartz_2020, Oberg_2016}. The subsequent radical formation triggers an exotic solid-state chemistry that results, among others, in the formation of larger, saturated species that play a role as the building stones of prebiotic species, i.e., sugars, fatty acids, alcohols, and even amino acids \cite{Ioppolo_Fedoseev_Chuang_Cuppen_Clements_Jin_Garrod_Qasim_Kofman_van_Dishoeck_etal_2021, Krasnokutski_2022, Holtom_2005, Jin_Garrod_2020}. To identify these species, and to characterize the reaction steps involved sophisticated spectroscopic and mass spectrometric detection techniques are needed, such as Fourier-transform infrared (FTIR) spectroscopy, reflection absorption infrared spectroscopy (RAIRS), or quadrupole mass spectrometry in combination with temperature-programmed desorption (QMS-TPD). 
 IR spectroscopy has the advantage that it can be used to identify the composition of ice and monitor in-situ the evolution of newly forming (or decomposing) species upon irradiation in a non-destructive manner. IR spectroscopy also has a few limitations; IR inactive modes cannot be measured, making some species such as O$_2$ or N$_2$ not observable. In the solid state, absorption bands are generally rather broad, and as a consequence, the spectral features of molecules with similar chemical functional groups tend to overlap, causing ambiguity. QMS-TPD is more sensitive than FTIR-RAIRS and can overcome some of the disadvantages but comes with the intrinsic limitation that it is a destructive method: to work, the ice has to desorb thermally. Moreover, TPD is limited to analyzing an ice composition only at its desorption temperature, which may differ from the composition at low temperatures at the end of irradiation \cite{Oberg_2016}. In the laboratory, a combination of both methods can be used to offer even more diagnostic potential through the use of molecular isotopes. However, both methods share the same shortcoming: they are not suited to investigate larger, more complex species in the ice. The IR spectra get too complex (start overlapping or lose intensity), the mass spectra lack sensitivity because of increasing numbers of fragment masses, moreover, the thermal desorption temperatures of the larger COMs are typically higher than at room temperature, which requires adapted cryostats. It is, in principle possible, to use more sensitive ex-situ mass spectrometric techniques such as gas or liquid chromatography, that can analyze the residue left on the substrate after a long duration of processing, but again, this does not allow in-situ investigations. Further uncertainties inherent to this method originate from heating samples up to room temperature, environmental exposure, chemical reactions with solvents, etc \cite{Oberg_2016}. 
 More recently, techniques such as time-of-flight mass spectrometry (ToF - MS) have been utilized in combination with other techniques to overcome these limitations and to in-situ characterize the constituents of the processed ices. For example, at W.M. Keck Research Laboratory in Astrochemistry at the University of Hawaii, ToF - MS driven by TPD is combined with single photon ionization, UV/Vis, and IR spectroscopy \cite{turner_exploiting_2020, jones_application_2013}. Another setup, the 2D-MALDI- ToF - MS experiment at the Ice Spectroscopy Lab (ISL) at Jet Propulsion Laboratory, NASA \cite{gudipati_situ_2012, yang_novel_2014} combines ToF - MS with laser ablation to study the constituents of processed ices without the need to interpret thermal ambiguities that may arise during TPD. This non-destructive way to probe the ice allows an in-situ analysis of the evolution of an interstellar ice analog upon (energetic) processing. In this approach, a generally non-reactive laser (low photon energy <30 mJ/cm$^2$) pulse is used to sublimate a small part of the irradiated ice. The resulting plume of molecules is ionized and then guided by mass selective ion optics onto a detector, in a manner that there is no contact between the extracted ions with other species or room-temperature components of the setup.
 
 Another setup and topic of this paper that utilizes a similar experimental methodology is the MATRI$^2$CES setup at Leiden University. In this setup, pulsed laser ablation is combined with ToF - MS, which comes with a high detection sensitivity \cite{Paardekooper_Bossa_Isokoski_Linnartz_2014}. Although powerful, these ToF - MS approaches are limited by the dynamic range of their detectors. Usually, the detector sensitivity is limited by the signal strength of the most abundant species. For signals originating from species at much lower abundance, this often prohibits unambiguous detections and simply increasing the detector sensitivity leads to signal saturation for the more abundant species which holds the potential to cause damage to the detector. 

 The focus of this article is to introduce an experimental concept to overcome this limitation by altering the ion beam traveling toward the detector so that the ions of the more abundant species are deflected and prevented from reaching the detector. This then allows for increasing the sensitivity of the detector without the risk of causing detector damage and allowing for the detection of less abundant species. This methodology is explained in the coming sections and is demonstrated by the use of argon and VUV irradiated methanol ices. 
 This method not only allows us to assign previously detected features of heavier species that have relatively low abundances with respect to the parent species more confidently but also allows us to see new, otherwise unobservable, species. As a proof of concept, in the next section results are presented for a pure argon ice, and the increase in the sensitivity is quantitatively discussed. Following this, the same technique is applied to methanol ice before and after VUV irradiation and the results with the new technique are compared with earlier findings on MATRI$^2$CES. The relevance for astronomical applications is briefly discussed.

\section{\label{sec:level1} Experimental Methodology: \protect}

 A detailed description of the construction, operation, data acquisition, and interpretation of MATRI$^2$CES can be found in previous works \cite{Paardekooper_Bossa_Isokoski_Linnartz_2014, Bulak_Paardekooper_Fedoseev_Linnartz_2021}. In the following section, a brief summary is presented followed by details of the new addition to the setup. 

\subsection{\label{sec:level2}Setup Description: }

 MATRI$^2$CES consists of two UHV chambers, the main chamber, and the time-of-flight mass spectrometer chamber, which are separated by a UHV gate valve. Both chambers are individually evacuated using turbo-molecular pumps and operate at a base pressure in the mid 10$^{-10}$ mbar range. 
 The main chamber (see also Figure 1) houses a gold-coated copper block that acts as a chemically neutral substrate on which ice is grown using a high-precision leak valve. The gold-coated substrate is attached to a closed-cycle helium cryostat that can cool the substrate down to 20 K. The temperature of the substrate is monitored using a silicon diode (Lakeshore DT-670-CU) that is connected to a temperature controller (Lakeshore 331).
 The temperature of the substrate can be regulated between 15 K and 300 K using resistive heating in thermal contact with the substrate. The cryostat is further attached to a two-dimensional translation module that allows the substrate to move along the horizontal (manually) and vertical (motorized) axes. 
 
 \begin{figure*}[htp]
 \includegraphics[width=\linewidth]{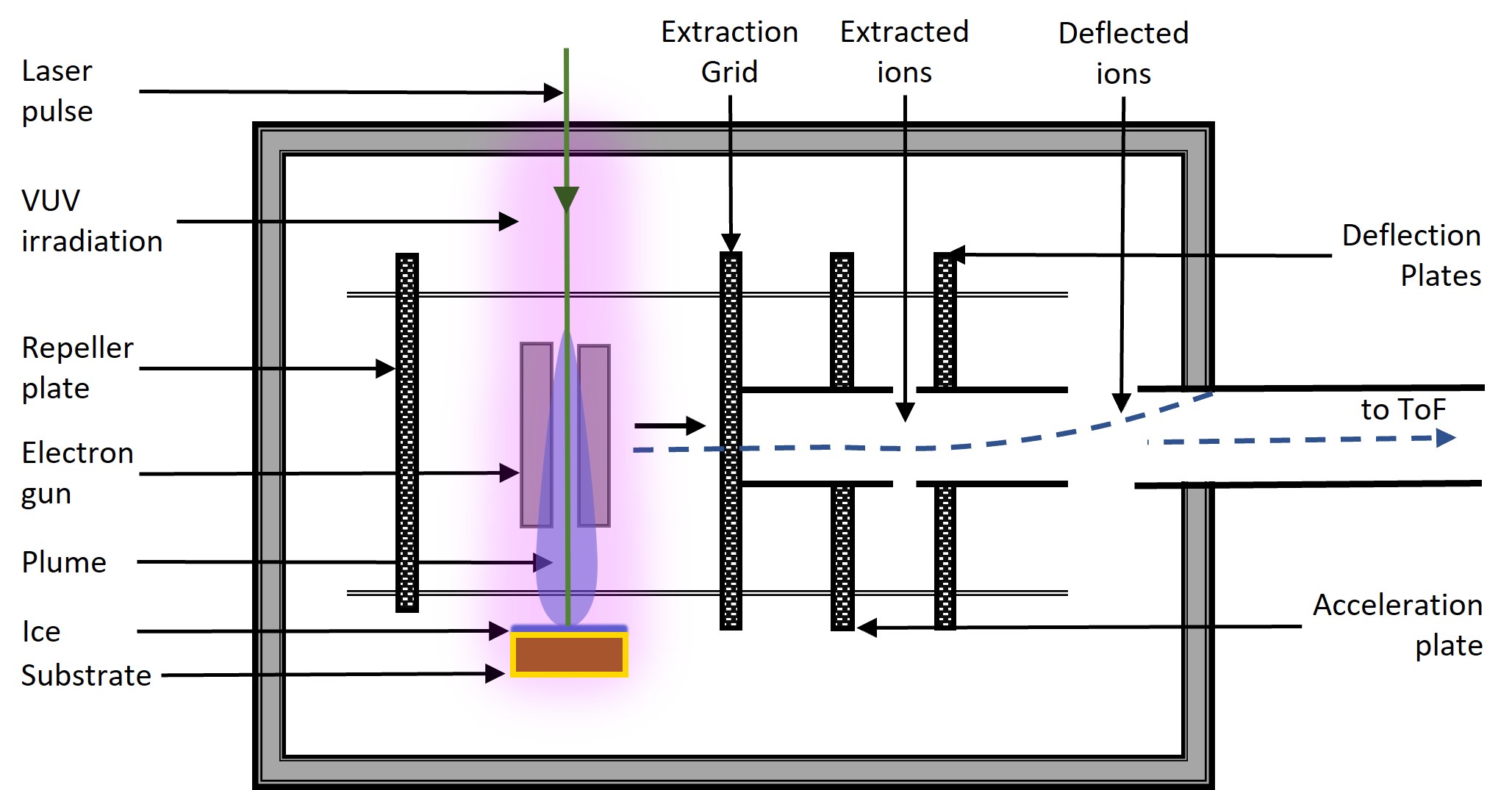}
 \caption{This figure illustrates the processing and extraction mechanism of ice constituents within MATRI$^2$CES. The ice is grown on a gold-plated copper substrate which is processed with VUV light. The (photo-processed) ice is then ablated using a laser pulse forming a gaseous plume of the resulting ice constituents, which are immediately ionized using an electron gun. The generated ions are then extracted by manipulating the voltage across the ion optics and accelerated towards the detector in the ToF - MS chamber. }
 \label{Setup}
\end{figure*}

 Ices grown on the substrate are processed by vacuum VUV photons produced using a Microwave Discharge Hydrogen Lamp (MDHL). The MDHL emits UV photons at a flux of (3 ± 0.5) x $10^{14}$ photons cm$^{-2}$s$^{-1}$ with a spectral energy distribution (SED) between 121.6 nm (10.2 eV) to 170 nm (7.2 eV), which is representative of the Lyman-alpha (121.6 nm) and the molecular hydrogen emission (130 - 165 nm) found in the ISM\cite{Prasad_Tarafdar_1983, Ligterink_Paardekooper_Chuang_Both_CruzDiaz_Van_Helden_Linnartz_2015}. The MDHL is attached directly in front of the substrate via a magnesium fluoride (MgF$_2$) view port that allows for uniform and complete VUV processing of the ice along the length and width of the substrate to be probed.

 Once UV photo-processed, the ice constituents are probed using laser desorption post-ionization time of flight mass spectrometry (LDPI ToF - MS). Within this scheme, the processed ice is ablated into the gas phase by using a short pulse of an unfocused aperture-adjusted Nd: YAG laser (New Wave Research, Polaris 2) operating in the third harmonic (355nm). The laser diameter is aperture adjusted to roughly 1.5 mm and the laser pulse duration is typically around 4-5 ns. The energy of the laser is controlled by an attenuator and optimized for a specific ice species in an energy range between 10 mJ cm$^{-2}$ to 50 mJ cm$^{-2}$ per pulse. This ensures minimal chemical impact due to the laser energy while aiming for complete ablation of the ice. 
 The laser pulse causes sharp local heating at the point of impact with the substrate, causing the ice to ablate from the substrate into a short-lived (µs) plume of the ice constituents in the gaseous state \cite{henderson_plume_2014}. This plume is formed orthogonally in front of the substrate. 

 After the ablation of the ice, the resulting plume constituents are immediately ionized using a continuously operating electron gun (Jordan C-950) which is placed directly below the expected plume position in front of the substrate. The electron gun generates electrons with an average electron energy of 70 eV, maximizing the ion yield. However, the high energy ionization also leads to fragmentation and (protonated) clusters that complicate the analysis and identification of newly formed species. Fortunately, the same species are also formed for unprocessed ices, which allows to discriminate between species formed upon VUV photolysis and ablation. Moreover, fragmentation patterns of the most common species are known from the literature, but in the case, that different species result in identical mass over charge values, ambiguity can arise. In such a situation, isotopically labeled precursors may help. 
 
 The generated ions are subsequently extracted using a set of ion optics. A repeller plate and an extraction grid are installed on either side of the region where the ions are generated. The repeller plate and the extraction grid are kept at the same potential (1050 V) during ion generation. After the ions are formed, the potential of the extraction grid is temporarily (4 µs) dropped by 100 V, causing ions to be extracted and further accelerated by the acceleration plate with constant potential, orthogonally toward a field-free drift tube in the second chamber, where they are detected by a micro-channel plate (MCP) detector placed inside the ToF - MS chamber. 
 
 Furthermore, since the ToF - MS is operating in reflection mode, the ions are gently deflected on the horizontal axis by applying a constant potential difference (100 V) to the deflection plates before the ions enter the ToF - MS chamber. The potential difference applied to these two deflection plates can be further manipulated.
\begin{figure}[htp]
 \centering
\includegraphics[width=0.8\linewidth]{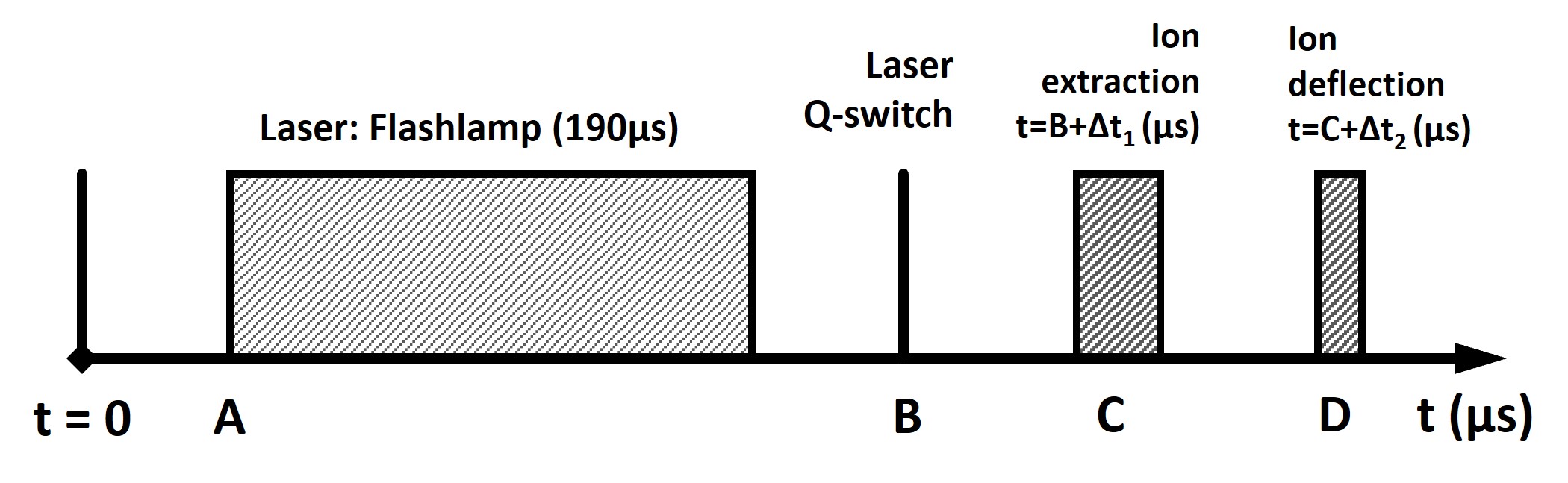}
 \caption{The time-sensitve operational sequence of LDPI- ToF - MS. The time from A to B represents the laser pulse on the substrate. After time C (B+$\bm{\Delta}$t$_1$) the generated ions are extracted by applying a potential difference to the extraction grid. After time D (C+$\bm{\Delta}$t$_2$), a short ($\bm{\mu}$s) potential difference is applied to alter the constituents of the beam of ions that enter the ToF - MS chamber.}
 \label{Timing}
\end{figure}
 The generation and extraction of ions is controlled by a series of precise micro-second adjustments and this is realized by using a pulsed delay generator (DG535 Stanford Research Systems). A typical timing scheme of an ion generation/extraction and detection cycle is given in Figure 2. The signal received by the MCP is recorded with a data acquisition card (DAQ) which collects data at a sampling rate of 2.5 x 10$^8$ Hz.

 For a typical experiment, a mass spectrum is obtained as the average of 100 individual measurements by probing 100 fresh spots on the substrate along a vertical column. This is done by coordinating the Nd:YAG laser pulse rate with the motorized vertical movement speed of the substrate. Averaging the 100 scans significantly decreases the overall noise in the final mass spectrum. This extraction cycle can then be repeated on a fresh vertical column by manually translating the substrate along the horizontal axes such that an averaged mass spectra can be acquired after a different dosage of VUV photons. This can be done for a total of 9 vertical columns. 

 For our experiments, we deposited three different molecules in our setup, Ar, $^{12}$CH$_3$OH and $^{13}$CH$_3$OH. In the case of argon, a 17.5 monolayers (ML) thick ice of pure Ar (Linde, 99.9999\%) was grown, and in the case of $^{12}$C- and $^{13}$C- methanol, 60 ML thick ices were grown. The methanol samples were put through 4 freeze pump thaw cycles to ensure the purity of the sample. While deposition, the thickness of the ice on the substrate was monitored using laser interference using a frequency-stabilized He-Ne laser which is placed at an angle of 2 degrees with respect to the normal of the substrate. The laser beam strikes the substrate and is reflected into a photo-diode where the beam is collected and converted into a digital signal \cite{Hudgins_Sandford_Allamandola_Tielens_1993}. 
 The thickness (d) of the ice grown is given by: 

\begin{equation}
d = \left(\frac{m\lambda}{2n_{\text{ice}} \cos\theta_{\text{ice}}}\right)
\label{eq:equation_label}
\end{equation}

where, m is the number of fringes, $\lambda$ is the wavelength of the laser beam (632.8 nm in case of He-Ne laser), n$_{ice}$ is the refractive index of the ice at a particular temperature, for argon ice \cite{Grace_Butcher_Monroe_Nikkel_2017} at 20 K n$_{ice}$ = 1.301 and for methanol ice \cite{Luna_Molpeceres_Ortigoso_Satorre_Domingo_Mate_2018} at 20 K n$_{ice}$ = 1.26, and $\theta$$_{ice}$ is the angle of reflection.

\vspace{0cm}
\subsection{\label{sec:level2}Upgrade and improvements. }

 The ability of MATRI$^2$CES to detect species with a very weak signal is primarily limited by the signal saturation of the MCP detector. This means that the strongest signal is case-determining. Quite often this is a signal directly related to a precursor species and not the resulting reaction products. To better detect such products that typically come in (much) lower abundances, the sensitivity of the MCP detector needs to be increased by applying a higher gain voltage. However, the increase of gain voltage that needs to be applied is restricted by the dynamic range of the MCP detector. In MATRI$^2$CES, a voltage of around 2500 V across the MCP is enough to saturate the detector signal (amplitude ~ 1 V). This voltage limit restricts the signal strength from less abundant species and the saturation comes with non-linearities. Repetitive measurements with signal saturation ultimately lead to detector damage resulting in sensitivity loss. Hence, to access the full potential of MATRI$^2$CES (and other systems using a similar concept), allowing for the detection of molecules with low abundances, it is imperative to increase the voltage gain applied across the MCP detector, but also ensure that species with high abundances do not saturate the detector. 

 This is done by introducing a second high-voltage pulser in the measurement scheme, as shown in Figure 2 (time D). This pulser supplies a precisely timed ($\mu$s) pulse of 50 V using a second delay generator (DG535 Stanford Research Systems) to the deflection plate. The controlled pulsed ion deflection (PID) allows for deflecting certain ions in a controlled mass range from traveling from the main chamber to the ToF - MS chamber. In order to probe the less abundant signals, we deflect ions that belong to the most abundant species to prevent them from reaching the detector. As a consequence of this PID, the MCP bias voltage applied to the detector can now be increased to enhance the sensitivity along the entire m/z range without saturating the detector. This results in a mass spectrum for which the dynamic range can be tailored to the strongest of the weaker mass signals. For our current experimental settings, this allows us to increase the MCP voltage settings from 2500 V to up to 3000 V. 

\section{\label{sec:level1} Results and Discussion\protect}
\subsection{\label{sec:level2}Argon Matrix }
 Argon was chosen to demonstrate the application of the PID method as the mass spectrum of Argon is fairly straightforward with little to no fragmentation, consisting of clear features of its three natural isotopes at m/z = 40 ($^{40}$Ar$^{+}$, 99.603\%), m/z = 36 ($^{36}$Ar$^{+}$, 0.334\%) and m/z = 38 ($^{38}$Ar$^{+}$, 0.063\%), along with peaks at m/z = 20 ($^{40}$Ar$^{2+}$), and m/z = 80 for $^{40}$Ar$_2$$^+$. 

\begin{figure}[htp!]
 \centering
\includegraphics[width=0.8\linewidth]{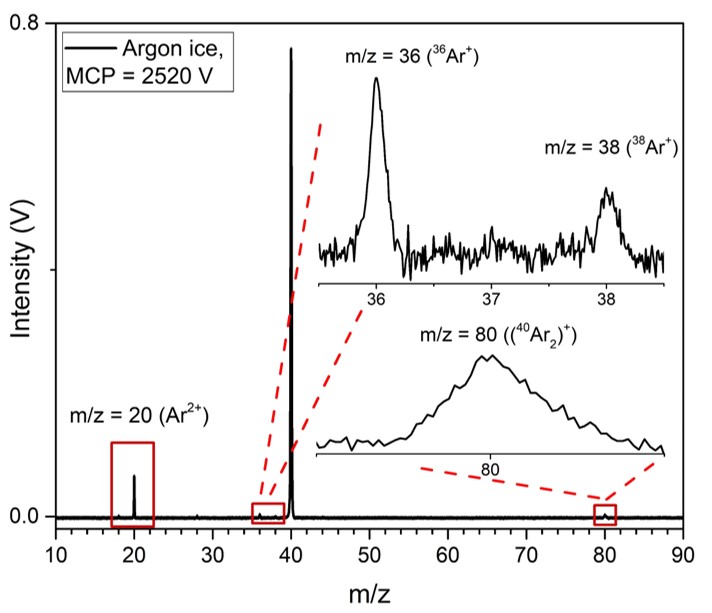}
 \caption{Mass spectrum obtained from the LDPI-ToF of 17.5 ML of an argon matrix at 20 K using a 70 eV ionization source.}
 \label{Ar}
\end{figure}

\begin{figure*}[htp!]
\includegraphics[width=1.2\linewidth]{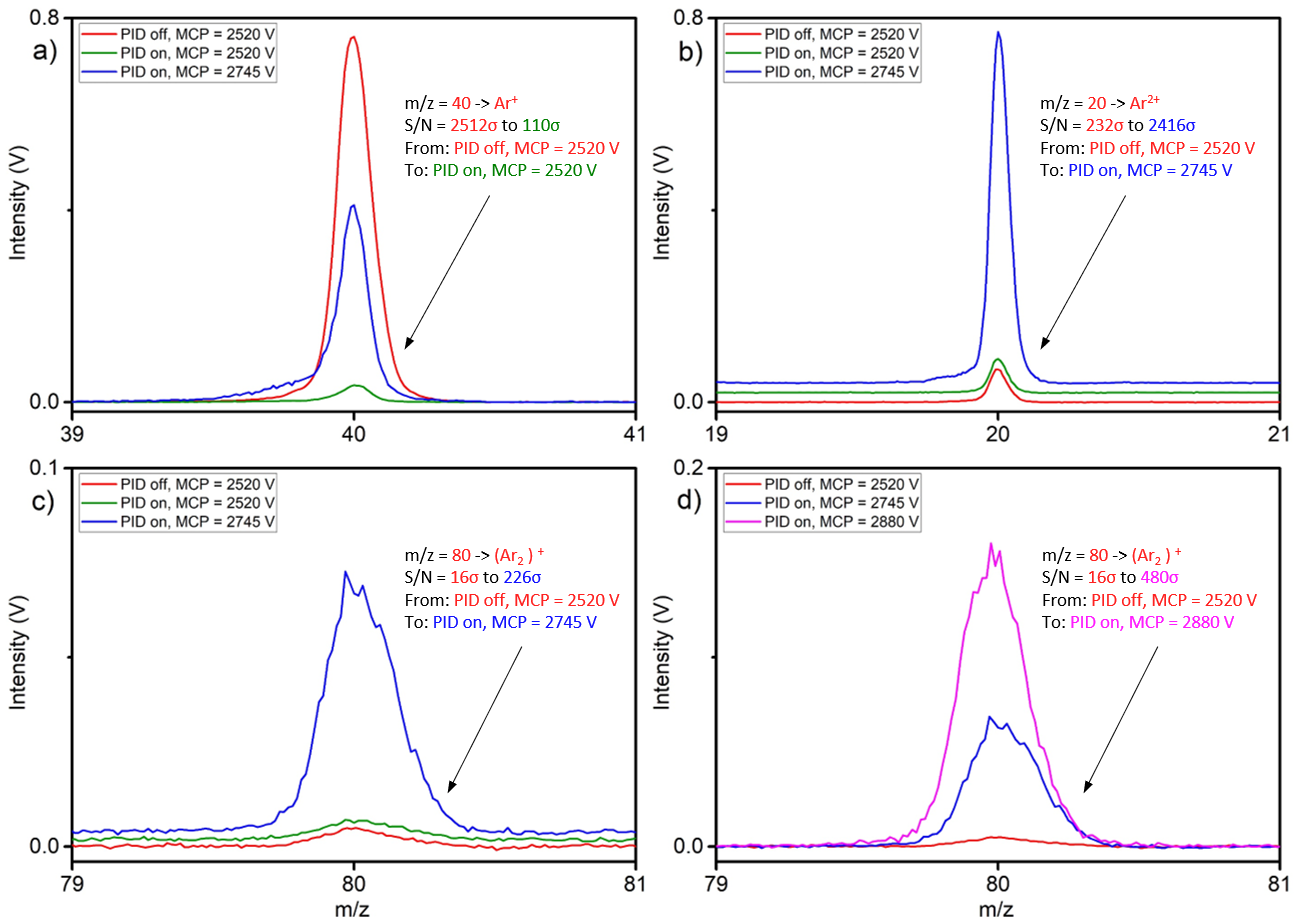}
 \caption{ Mass spectra are shown for ToF - MS of argon features at m/z = 40 (Ar$^+$), m/z = 20 (Ar$^{2+}$), and m/z = 80 ((Ar$_2$)$^+$) in panels (a), (b), and (c), respectively. The MS is obtained at an MCP bias voltage of 2520 V without PID (Red), MCP bias voltage of 2520 V with PID (Green), and an MCP bias voltage of 2745 V with PID (Blue). Panel (d) shows MS for m/z = 80 ((Ar$_2$)$^+$) at three different MCP bias voltages of 2520 V (Red), 2745 V (Blue), and 2880 V (Pink). MS in panels (b) and (c) are offset for clarity.}
 \label{Argon P2}
\end{figure*}

 Since the more abundant peaks of argon at m/z = 20, 40, and 80 are separated by relatively large m/z values, it is an ideal element to illustrate the basic concept of the two-step PID method. To qualitatively compare the detection sensitivity of the ToF - MS with and without the application of the second pulser, we use the signal-to-noise ratio (S/N, $\bm{\sigma}$), which is defined as the ratio of the measured signal with the background noise. The $\bm{\sigma}$ values are calculated by dividing the maximum peak intensity of a specific mass signal by the standard deviation of the noise signals along a wide m/z range where we do not detect any other mass peaks. A minimum of three times the S/N ratio (3$\bm{\sigma}$) is required for successful detections and is defined as the Limit of Detection (LOD). 
 Argon was deposited at 20 K and its ToF - MS is shown in Figure 3. The bias voltage applied to the MCP is 2520 V, and in the MS we can see that the peak intensity at m/z = 40 is at 0.8 V, which is close to saturation (amplitude = 1 V), while the peak at m/z = 20 is significantly smaller at 0.06 V and the peak at m/z = 80 is barely seen at 0.005 V. Here the signal of m/z = 36, $^{36}$Ar$^{+}$ corresponds to 0.334\% of the total argon signal and the detector sensitivity at MCP = 2520 V corresponds to a 7 $\bm{\sigma}$ detection. Considering that 1 ML of ice corresponds to 10$^{15}$ molecules/cm$^2$, this gives us a base sensitivity of m/z = 36 at 7$\bm{\sigma}$ of 5.84 x 10$^{13}$ molecules/cm$^2$, which would correlate to a 3$\bm{\sigma}$ value of 2.5 x 10$^{13}$ molecules/cm$^2$. This puts the LoD without using PID at 2.5 x 10$^{-2}$ ML. 

 To overcome the limitation of signal saturation due to the m/z = 40 peak, we employ the PID technique to selectively deflect the ions in the unwanted mass range. For this, a second pulser applies a short (0.5 $\mu$s) pulse after a calibrated time interval $\bm{\Delta}$t$_2$ of 1.95 $\mu$s to drop the voltage across the deflection plates by 50 V. These values are obtained after a series of optimization attempts wherein the trigger time and the duration of the pulse are varied and their impact on the resulting mass spectrum is noted. Once the signal at m/z = 40 is reduced by deflecting $^{40}$Ar$^+$ from the path towards the detector (though not fully eliminated) the voltage applied to the MCP detector is now increased until the detector is close to saturation by the second most intense peak of m/z = 20, which in our case is 2745 V, This leads to improved detector sensitivity for other mass ranges in the spectrum. Figure 4 gives an overview of the two-step technique resulting in changes for features at m/z = 40, 20, and 80 in panels 4(a), 4(b), and 4(c), respectively, while panel 4(d) shows further enhancement in the signal-to-noise ratio by increasing the MCP voltage further to 2880 V. In Figures 4(a) to 4(c), the red spectra correspond to the mass spectra of argon without the use of PID with a bias MCP voltage = 2520 V. The green peak corresponds to mass spectra while using PID at an unchanged MCP bias voltage of 2520 V. The blue peak corresponds to mass spectra with PID at an increased MCP bias voltage of 2745V. Figure 4(d) shows the mass spectra of m/z = 80 at three different MCP values: 2520 V/pulser off (Red), 2745 V, pulser on (Blue), and 2880 V, pulser on (Pink). 

 In panel 4(a), for the setting without using PID at an MCP bias of 2520 V the intensity of the m/z = 40 feature is at 0.8 V. After the application of PID, without changing the MCP bias voltage, the value decreases to 0.03 V, as shown by the green peak. For a higher MCP bias voltage using PID, the signal intensity increases again as shown by the blue peak, improving the detection sensitivity in this mass range. In panels 4(b) and 4(c) the signals with PID and without PID at an MCP bias of 2520 V are similar as expected since the PID is only utilized for the m/z = 40 feature, however, upon increasing the MCP bias to 2745 V, the intensity of the m/z = 20 peak increases from 0.06 V to 0.8 V and m/z = 80 peak increases from 0.005 V to 0.08 V. A stronger signal is realized only for the ‘pulser on’ setting.
 It should be noted that the red and green spectra of m/z = 20 and 80 in panels 4(a) and 4(c) remain unaffected upon the application of the pulser, without changing the bias voltage across the MCP. This emphasizes the fact that PID does not interfere with the signal intensities of features outside the m/z range that is targeted for deflection. 

 The dampening of the signal intensity of the more abundant feature allows us to increase the MCP bias voltage until the ToF - MS is saturated by the next most abundant species, in this case, the blue spectra in panel 4(b) at m/z = 20. Here, the S/N of the feature has increased from 232$\bm{\sigma}$ to 2416$\bm{\sigma}$. At this setting, the feature at m/z = 80 is also shown in panel 4(c), and the S/N value can be seen to increase from 16$\bm{\sigma}$ to 226$\bm{\sigma}$. This increase in sensitivity is limited by the ions with m/z = 20. To overcome this secondary signal saturation event and further enhance the detection capabilities of the setup, it is possible to increase the pulse width and deflect not only m/z = 40 but also the ions of m/z = 20 which further increases the detection sensitivity of features with larger m/z values (see panel 4(d)). 
 Consequently, the pulse duration of the second pulse is adjusted from 0.5µs to 2µs so that all signal below m/z = 40 is deflected. This deflection of the m/z = 20 feature allows us to further increase the MCP bias voltage to 2880V as shown in panel 4(d), during which the intensity of the m/z = 80 peak further increases from 0.08 V to 0.16 V, resulting in a S/N ratio increase to 480$\bm{\sigma}$, which is 30 times higher than the original signal. Under the assumption that the increase in the intensity of the features in the mass range of m/z = 20 to 80 is similar, we expect the new LoD to be 8 x 10$^{11}$ molecules/cm$^2$.

 The intensity of the m/z = 80 feature at an MCP bias voltage = 2880 V is at 0.175 V. The maximum bias voltage we can apply to our MCP is restricted to 3500 V, which in principle leaves space for further improvement by simply increasing the MCP voltage. In the present system, the use of bias MCP voltages close to 3500 V can cause an increase in the noise captured in the spectrum, negatively impacting the S/N ratio. It is not possible to employ PID in a manner that we only dampen the signal of the m/z = 40 feature and not of m/z = 36 or 38 features, which becomes visible when also comparing the mass signals for $^{40}$Ar$^+$, $^{36}$Ar$^+$ and $^{38}$Ar$^+$. At present, the smallest mass range that can be dampened without affecting the features around the deflection zone is approximately 10 atomic mass units (amu).

\subsection{\label{sec:level2} Methanol Ice }

 The same two-step PID technique is applied to study astronomically relevant methanol ice, both before and after VUV irradiation. Methanol is considered a very important molecule in the formation of complex organic molecules (COMs) in the ISM. The abundance of methanol ice, with respect to water, has been derived between 5-12\% in dense clouds and 1-30\% in cold regions around low and high-mass protostars\cite{boogert_observations_2015, Gibb_Whittet_Boogert_Tielens_2004, dartois_methanol_1999, pontoppidan_detection_2003}.
 Methanol ice has been studied in the laboratory in various interstellar environments \cite{cruz_2016, Andrade_2009, Almeida_2012, Dartois_2020}; methanol has been shown to interact with energetic and non-energetic processes \cite{chen_soft_2013, mathew_methanol_2022, abou_mrad_methanol_2016, freitas_laboratory_2020, Munoz_Caro_Ciaravella_2019} and to actively take part in interstellar chemistry when mixed with other ices\cite{Mason_2014, maity_infrared_2014, maity_formation_2015, de_Marcellus_2015, Bernstein_1995}.
 
 \begin{figure}[]
\includegraphics[width=0.6\linewidth]{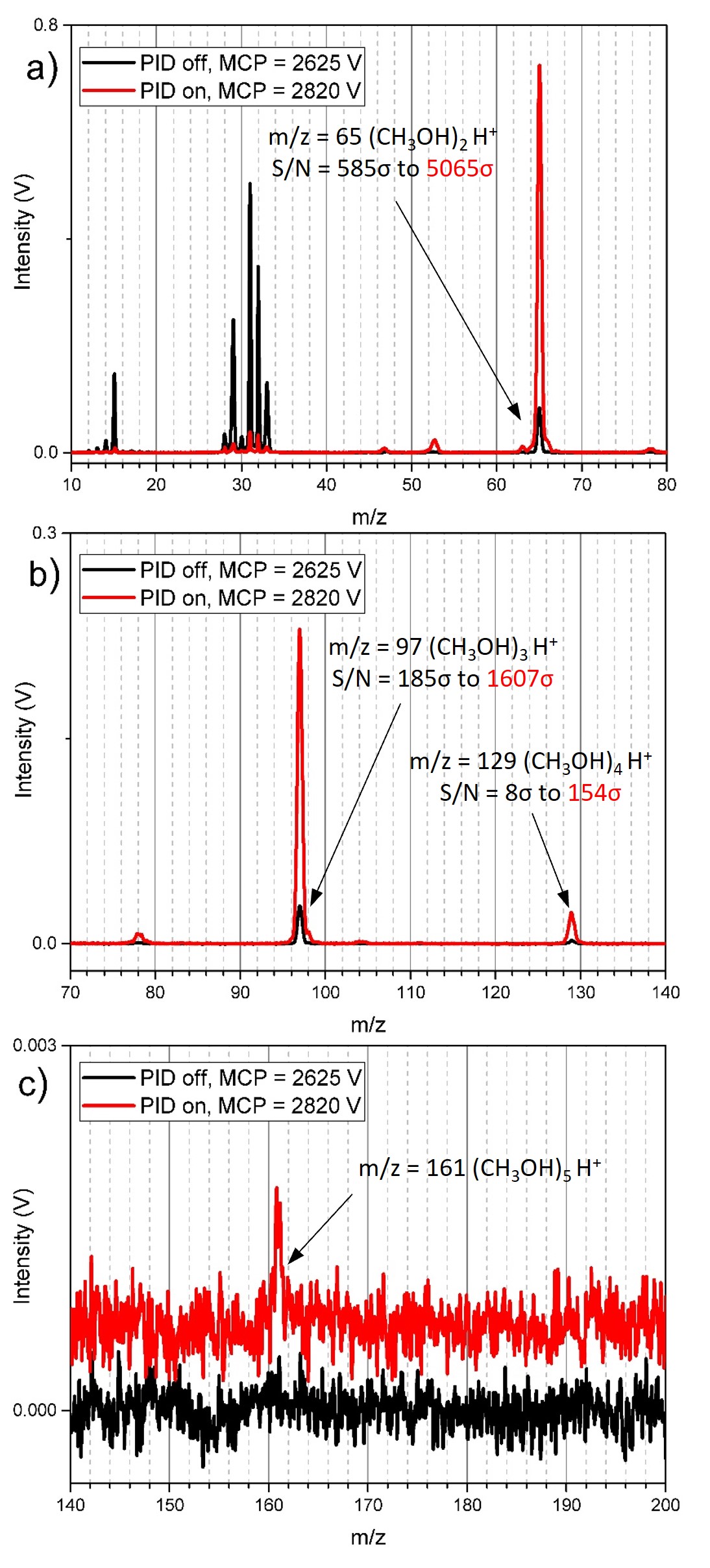}
 \caption{The LDPI ToF – MS of non-UV irradiated pure methanol ice is shown without (2625 V) and with (2820 V) applying the two-step PID technique. Panels (a), (b), and (c) show mass spectra covering m/z = 0-70, 70-140 and 140-200, respectively. The increase in the S/N ratios becomes particularly visible by looking at protonated methanol clusters ((CH$_3$OH)$_n$H$^+$) at m/z = 65, 97, and 129 for n = 2,3, and 4, alongside the previously undetected signal at m/z = 161 for n = 5.}
 \label{Methanol P2}
\end{figure}

 \begin{figure*}[]
\includegraphics[width=1.15\linewidth]{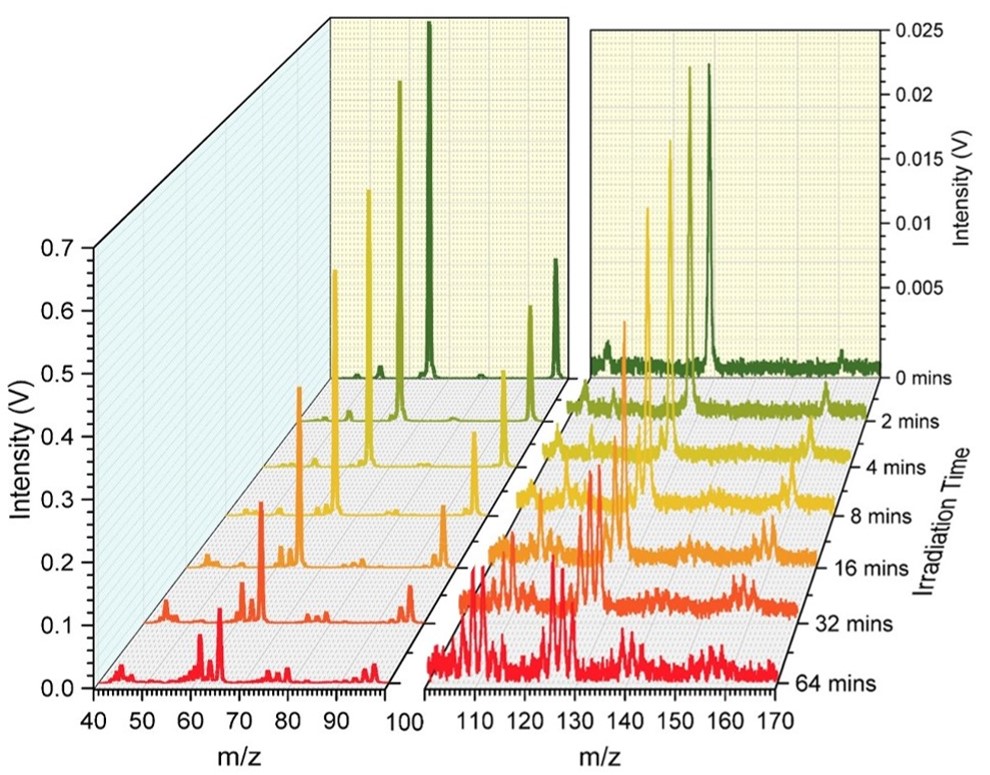}
  \caption{Progression of the post-PID methanol ice mass spectra, at MCP = 2820 V, with different VUV exposure times, ranging from zero to approximately 1.8x10$^{18}$ photons/cm$^2$ of VUV fluence, for mass ranges between m/z = 40-100 and m/z = 100-170. Note the different vertical axes (intensity).}
  \label{methanol progression}
\end{figure*}

 In earlier works on MATRI$^2$CES\cite{paardekooper_laser_2016}, the formation of a wealth of COMs was shown upon VUV processing of methanol ice. COMs were identified with the help of reference experiments that provided data on the thermal desorption temperatures of the molecules. A wealth of COMs were reported, including formaldehyde (H$_2$CO), dimethyl ether (CH$_3$OCH$_3$), acetaldehyde (CH$_3$CHO), methyl formate (HCOOCH$_3$), ethanol (CH$_3$CH$_2$OH), formic acid (HCOOH), acetic acid (CH$_3$COOH), glycolaldehyde (HOCH$_2$CHO), methoxy-methanol (CH$_2$OCH$_2$OH), and ethylene glycol (CH$_2$OH)$_2$ by analyzing the mass spectra features until m/z = 62 \cite{paardekooper_laser_2016}. The largest COM suggested in this study was glycerine (glycerol - C$_3$H$_8$O$_3$) based on the TPD residue, while signals were detected until m/z = 97 corresponding with the formation of a protonated methanol cluster (more in the next section). Many of these species meanwhile have been identified in radio astronomical surveys \cite{McGuire_2022}. In a different, more recent study using GCMS as an ex-situ diagnostic tool for a UV-irradiated methanol ice sample, even heavier COMs such as methyl propionate (CH$_3$CH$_2$CO$_2$CH$_3$) and methyl butyrate (CH$_3$CH$_2$CH$_2$CO$_2$CH$_3$) have been reported \cite{tenelanda-osorio_effect_2022}.
  
 In our work, methanol ice was deposited at 20 K and the resulting mass spectra of pure methanol ice, with and without the PID technique, are shown in Figure 5. As demonstrated with the argon example, we aim here to avoid the strong signals originating from the parent (CH$_3$OH) species between m/z = 27 and m/z = 33. We found that after using the PID technique to reduce the methanol signal, the signals of the major methanol fragments (upon electron impact ionization) between m/z = 12 and 19 quickly saturated the detector. To suppress these, a longer pulse that deflects not only the methanol parent peak but also the signal of its major fragments was needed, the PID was set for a pulse with a duration of 1.5 µs and at a time t$_2$ = 4 µs after the extraction (see Figure 2), resulting in a mass spectrum with enhanced signal intensity for masses above m/z = 40. 

 The ToF - MS of pure non-UV irradiated methanol ice (black curve in panel 5(a)) was recorded at an MCP bias of 2625 V, showing the main fragmentation peaks at m/z = 12 (C$^+$), 13 (CH$^+$), 14 (CH$_2$$^+$), 15 (CH$_3$$^+$), 16 (CH$_4$$^+$), 17 (OH$^+$), 18 (H$_2$O$^+$) and 19 ((H$_2$O)H$^+$), 28 (CO$^+$), 29 (CHO$^+$), 30 (CHOH$^+$), 31 (CH$_2$OH$^+$), 32 (CH$_3$OH$^+$) and 33 ((CH$_3$OH)H$^+$). The ToF - MS also shows features that correspond to protonated methanol clusters in the form of (CH$_3$OH)$_n$H$^+$ at n = 1, 2, 3, and 4 corresponding to m/z = 33, 65, 97, and 129, respectively (see also panel 5(b)). These clusters are found both in UV and non-UV irradiated experiments and are formed during the laser ablation process via a clustering reaction represented in the following equation \cite{Focsa_Mihesan_Ziskind_Chazallon_Therssen_Desgroux_Destombes_2006}.
 \begin{equation}
 (CH_3OH)_{n+1}^+ \rightarrow (CH_3OH)_nH^+ + CH_3O
\label{}
\end{equation}

 These clusters have been reported in earlier works and are an intrinsic result of the detection method. They can introduce ambiguity for data interpretation, as the corresponding mass features can overlap with the newly formed reaction products upon UV irradiation. On a positive note, this also offers a very accurate mass calibration tool, and the ambiguities caused by these clusters can be circumvented to a great extent by repeating measurements on grown ices with a different isotopic composition. 

 Figure 5(a) (red) shows that the main methanol fragment signals below m/z = 33 are dampened because of the PID technique. This allows us to increase the sensitivity of the detector by increasing the bias voltage supplied to the detector to 2820 V until the peak of (CH$_3$OH)$_2$H$^+$ at m/z = 65 almost saturates the detector. The increase in the signal strength of protonated clusters (CH$_3$OH)$_3$H$^+$ at m/z = 97 and (CH$_3$OH)$_4$H$^+$ 129 can be seen in panel 5(b) and the previously undetected feature of (CH$_3$OH)$_5$H$^+$ at m/z = 161 shows up in panel 5(c). It is possible to further increase the pulse duration of the PID to deflect ions corresponding to m/z = 65 (the protonated methanol dimer), allowing to further increase the voltage settings, until the next cluster is case limiting, etc.

Here the PID is further optimized to study the photolysis products of VUV irradiated methanol ice. The mass spectrum is recorded starting from a VUV exposure of 2, 4, 8, 16, 32, and 64 minutes, corresponding to a total photon fluence of 1.8 x 10$^{18}$ photons cm$^{-2}$. The results are summarized in Figure 6 where post-PID MS of methanol ice are shown for mass ranges: m/z = 40 to 170. In the mass spectra, the formation of new larger species with increasing VUV fluence is evident, particularly, in the mass ranges m/z = 40 to 50, m/z = 70 to 80, m/z = 105 to 115, and m/z = 135 to 145. Not only the formation of photo-products can be seen, but also the destruction and/or consumption of species with mass signals at m/z = 52, 104, or at m/z = 65, 97, 129, 161, representative for the consumption of methanol, becomes visible.

 In Figure 7, the mass spectra of irradiated methanol ice with and without the PID technique are compared for mass ranges: m/z = 40 to 100 and m/z = 100 to 170 in panels 7(a) and 7(b), respectively. Due to the enhanced sensitivity, the "PID on" (red) spectrum has a wealth of new peaks not seen in the "PID off" spectrum. Notably, peaks in the mass ranges m/z = 50 to 60, m/z = 68 to 74, m/z = 80 to 92, m/z = 105 to 115, m/z = 135 to 145, and m/z = 153 to 161. 

 To investigate the molecular origin of these newly observed mass peaks upon VUV irradiation, the same experiment was repeated for a $^{13}$CH$_3$OH ice, and the resulting mass spectra with PID are compared with regular methanol ice in Figure 8. The use of a $^{13}$C isotopically enriched sample allows us to confidently assign the number of C atoms of a specific mass peak carrier, and to roughly determine the chemical formula (i.e., the composition of H, C, and O atoms) of the species. The corresponding assignments are shown in the figure. For example, the mass signals at m/z = 75, 77, and 79 have corresponding methanol-${13}$C mass signals at m/z = 78, 80, and 82, suggesting that the carrier species contains 3 carbon atoms, and consequently, it can have 2 oxygen atoms and 7, 9, and 11 hydrogen atoms respectively (i.e., C$_3$H$_{7-11}$O$_2$).
\begin{figure*}[]
\includegraphics[width=1.2\linewidth]{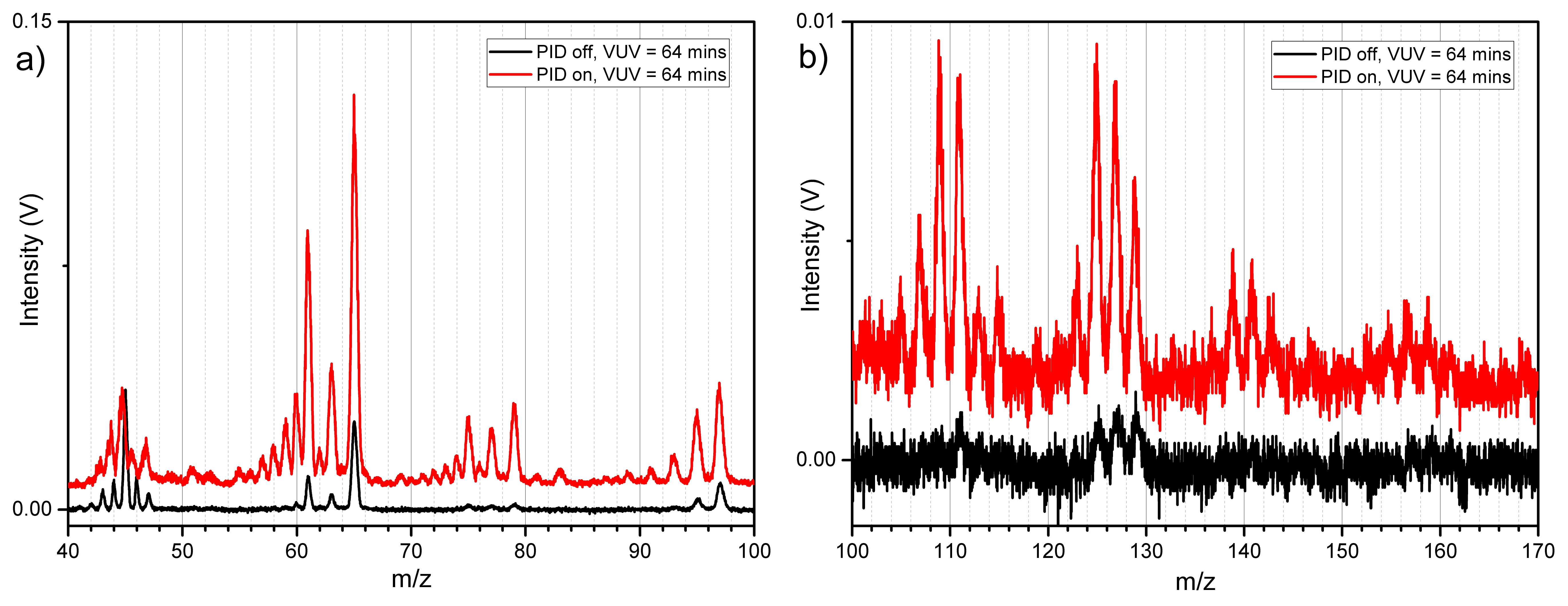}
 \caption{A comparison of with (Red) and without (Black) PID MS of pure irradiated methanol ice at m/z ranges between m/z = 40 and m/z = 100 and between m/z = 100 and m/z = 170, is shown in panels (a) and (b) respectively, with offset for clarity}
 \label{methanol 12 pid}
\end{figure*}

 A detailed analysis of these mass peaks is beyond the scope of this work. Many mass peaks may originate from mass fragments, similar mass peaks may originate from different precursors, and for one mass peak, different isomers may exist. We expect that with the help of complementary tools, such as TPD, the use of deuterated precursors and systematic variations of the laser ablation intensity and used electron impact ionization energies \cite{paardekooper_laser_2016}, it will be possible to disentangle the complex mass spectra. 
 \begin{figure*}[]
\includegraphics[width=1.2\linewidth]{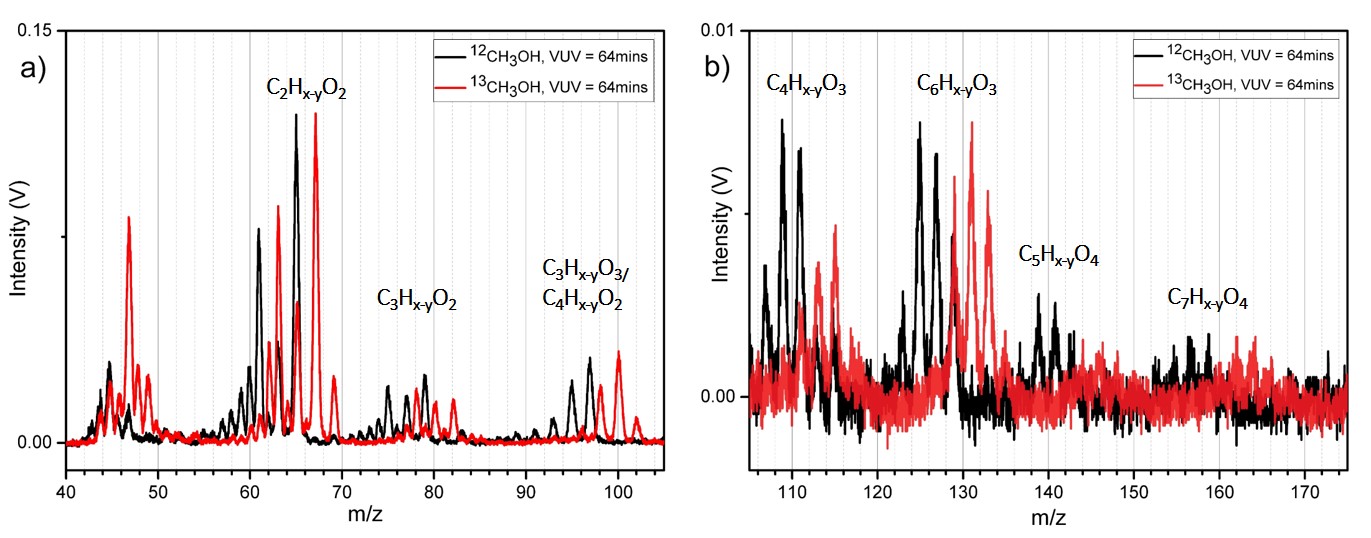}
 \caption{ A comparison of irradiated MS between 12C (Black) and 13C (Red) methanol for m/z ranges between m/z = 40 and m/z = 105 and between m/z = 105 and m/z = 175, is shown in panels (a) and (b) respectively.}
 \label{methanol 12 vs 13}
\end{figure*}
 The take-home message here focuses on an experimental concept showing that it is possible to record larger COMs by actively manipulating the generated ion beam and the detector saturation voltages. In comparison with the earlier study \cite{paardekooper_laser_2016}, the new ion-deflected mass spectrum shows a wealth of previously undetected features until m/z = 160 and makes it possible to identify much larger COMs that can be formed upon VUV irradiation of simple yet abundant ices in the ISM. In this work, the formation of photoproducts upon increasing VUV fluence is clearly seen within the mass range between m/z = 75 to m/z = 160. From an astronomical perspective, the results from our work also provide further evidence to conclusions from previous studies suggesting that methanol is a key precursor molecule in the formation of larger COMs upon VUV irradiation \cite{Oberg_Garrod_van_Dishoeck_Linnartz_2009, paardekooper_laser_2016}. Some of these COMs have already been identified in the gas phase\cite{VanGelder_Tabone_Tychoniec_Van_DIshoeck_Beuther_Boogert_Caratti_Garatti_Klaassen_Linnartz_Muller_etal_2020, Nazari_VanGelder_Van_Dishoeck_Tabone_VanT_Hoff_Ligterink_Beuther_Boogert_Caratti_Garatti_Klaassen_etal_2021} as they transfer from the solid state into the gas phase as the star formation cycle progresses (due to different desorption mechanisms that comprise reactive, photon and thermally induced desorption). This further highlights the importance of the PID detection technique as the larger COMs detected in our work are expected also to be present in the ISM, but in quantities that are beyond the current detection capabilities. The experimental findings also fill the gap between RAIRS/TPD data and the results from ex-situ methods such as GCMS that hint at the formation of even larger species \cite{Munoz_2002, Greenberg_Gillette_Caro_Mahajan_Zare_Li_Schutte_Groot_Mendoza_Gomez_2000}. It is likely that in more ISM representative ices, with other species (like water), this chemistry only gets more complex as shown for VUV irradiated CH$_3$CN and H$_2$O ice mixtures \cite{Bulak_Paardekooper_Fedoseev_2021} emphasizing the importance of the data that can be collected of irradiated mixed ices using the PID technique. The present version of MATRI\textsuperscript{2}CES holds the potential to be advanced with other mass segregation techniques, such as quadrupole mass filter (QMF) and ion trap, however, for now, this is beyond the current scope of this paper.

 To correctly identify photo-products at such high sensitivities, it is essential to distinguish between newly formed photo-products and protonated clusters formed upon last ablation. The intrinsic appearance of protonated cluster ions using this method further underlines the relevance of the method introduced here. A study using a similar experimental technique but without PID, studied the interaction of methanol ice with high energy electrons (2 KeV) and VUV (121.6 - 160 nm) photons \cite{henderson_direct_2015}. In their study with UV photons, mass peaks at m/z = 33, 65, 97, and 127 were directly assigned to protonated methanol clusters in the form of (CH$_3$OH)$_{1-4}$H$^+$. They also reported methanol clustering with its oxidation products, for example (CH$_3$OH)(H$_2$CO)H$^+$ at m/z = 63, (CH$_3$OH)$_2$(H$_2$CO)H$^+$ at m/z = 95, (CH$_3$OH)$_2$(H$_2$CO)$_2$H$^+$ at m/z = 125. Hence in our work m/z values of 61, 63, 91, 93, 95, 123, 125, 127, 155, 157, and 159 can also have contributions, but due to the close pattern of mass peaks of these clusters, it is easy to distinguish them. It is also possible to track the clustering of individual species along the whole m/z range, and by comparing regular methanol with the methanol-$^{13}$C results. We can determine the clusters from the products with the process of elimination, for example, the feature at m/z = 93 could correspond to an ethanol (photo-product of methanol) cluster\cite{henderson_direct_2015} (C$_2$H$_6$O)$_2$H$^+$. However, the methanol-$^{13}$C feature at m/z = 96 confirms the presence of 3 carbon atoms instead of 4. Hence the possibility of ethanol clusters can be eliminated in our case. Hence, the features between mass ranges m/z = 70 and 82, m/z = 106 and 114, and m/z = 136 and 144 exclusively correspond to C$_3$H$_{x-y}$O$_2$, C$_4$H$_{x-y}$O$_3$, and C$_5$H$_{x-y}$O$_4$, which are reported for the first time. It should also be noted that such strong clustering is not expected in non-polar molecules.  
 Despite these advantages, this technique has its own set of challenges. The analysis of the data in higher m/z values becomes exceedingly challenging as the fragmentation patterns of many of these species have not yet been documented in databases. The analysis of heavier photoproducts will also benefit from a library of reference measurements by depositing expected reaction products and specifically measuring their (UV irradiated) mass spectra. This is a work in progress that has become possible by introducing PID. 

\section{\label{sec:level1} Conclusions\protect}
 
 The present work shows that the use of the PID technique in a laser desorption post-ionization time-of-flight mass spectrometric system, optimized to study the formation (and destruction) of photoproducts formed in astronomically relevant ice analogs, allows for substantially increasing the detection limit of species that are present in lower abundances, typically at higher m/z values. The concept is based on sequentially decreasing the strongest mass signals by PID, allowing an increase of the detector voltage settings (sensitivity) without saturation to monitor species that are produced in lower abundances. This concept is demonstrated in the example of frozen Ar, showing an overall detection limit increase by a factor of 30. In the case of pure methanol ice, many new mass peaks can be seen. These peaks correspond to species not formed easily and are typically found in mass ranges inaccessible in earlier work. With this step realized, additional tools, such as isotopic labeling, can be used for further COM identifications. We expect this method to be a relatively cheap, beneficial addition for similar experimental setups with different applications as well.

\begin{acknowledgments}
This work has been supported by the Danish National Research Foundation
through the Center of Excellence InterCat (grant agreement No. DNRF150).
It has also been funded by the Dutch Astrochemistry Network II (DANII)
and NOVA, the Netherlands Research School for Astronomy.
\end{acknowledgments}
\section*{Conflict of Interest}

The authors have no conflicts to disclose.

\section*{Data Availability}

All data is stored on servers in-house. The data that support the findings of this study are available from the corresponding author upon reasonable request.

\bibliographystyle{aipnum4-1}
\bibliography{RSI}

@PREAMBLE{
 "\providecommand{\noopsort}[1]{}" 
 # "\providecommand{\singleletter}[1]{#1}%" 
}

@article{McGuire_2022, title={2021 Census of Interstellar, Circumstellar, Extragalactic, Protoplanetary Disk, and Exoplanetary Molecules}, volume={259}, number={2}, journal={The Astrophysical Journal Supplement Series}, author={McGuire, Brett A.}, year={2022}, pages={30} }

@article{belloche_re-exploring_2019,
	title = {Re-exploring {Molecular} {Complexity} with {ALMA} ({ReMoCA}): interstellar detection of urea},
	volume = {628},
	shorttitle = {Re-exploring {Molecular} {Complexity} with {ALMA} ({ReMoCA})},
	
	journal = {A\&A},
	author = {Belloche, A. and Garrod, R. T. and Müller, H. S. P. and Menten, K. M. and Medvedev, I. and Thomas, J. and Kisiel, Z.},
	  
	year = {2019},
	pages = {A10},
	
}

@article{guelin_organic_2022,
	title = {Organic {Molecules} in {Interstellar} {Space}: {Latest} {Advances}},
	volume = {9},
	shorttitle = {Organic {Molecules} in {Interstellar} {Space}},
	
	journal = {Frontiers in Astronomy and Space Sciences},
	author = {Gu{\'e}lin, Michel and Cernicharo, Jose},
	year = {2022},
}

@article{Yang_Green_Pontoppidan_Bergner_Cleeves_II_Garrod_Jin_Kim_Kim_etal_2022,
        title={Corinos. {I.} {JWST}/{MIRI} spectroscopy and imaging of a class 0 protostar {IRAS} 15398–3359}, 
        volume={941}, 
      
        author={Yang, Yao-Lun and Green, Joel D. and Pontoppidan, Klaus M. and Bergner, Jennifer B. and Cleeves, L. Ilsedore and Evans II, Neal J. and Garrod, Robin T. and Jin, Mihwa and Kim, Chul Hwan and Kim, Jaeyeong and Lee, Jeong-Eun and Sakai, Nami and Shingledecker, Christopher N. and Shope, Brielle and Tobin, John J. and Dishoeck, Ewine F. van},
        year={2022},
      
        journal = {Astrophys. J.},
        pages={L13},
    }

@article{McClure_Rocha_Pontoppidan_Crouzet_Chu_Dartois_Lamberts_Noble_Pendleton_Perotti_etal_2023,
title={An {Ice} {Age} {JWST} inventory of dense molecular cloud ices}, volume={7}, journal={Nature Astronomy}, author={McClure, M. K. and Rocha, W. R. M. and Pontoppidan, K. M. and Crouzet, N. and Chu, L. E. U. and Dartois, E. and Lamberts, T. and Noble, J. A. and Pendleton, Y. J. and Perotti, G. and Qasim, D. and Rachid, M. G. and Smith, Z. L. and Sun, Fengwu and Beck, Tracy L. and Boogert, A. C. A. and Brown, W. A. and Caselli, P. and Charnley, S. B. and Cuppen, Herma M. and Dickinson, H. and Drozdovskaya, M. N. and Egami, E. and Erkal, J. and Fraser, H. and Garrod, R. T. and Harsono, D. and Ioppolo, S. and Jiménez-Serra, I. and Jin, M. and Jørgensen, J. K. and Kristensen, L. E. and Lis, D. C. and McCoustra, M. R. S. and McGuire, Brett A. and Melnick, G. J. and Öberg, Karin I. and Palumbo, M. E. and Shimonishi, T. and Sturm, J. A. and van Dishoeck, E. F. and Linnartz, H.}, year={2023}, pages={431}, }

@article{Ioppolo_Fedoseev_Chuang_Cuppen_Clements_Jin_Garrod_Qasim_Kofman_van_Dishoeck_etal_2021,
title={A non-energetic mechanism for glycine formation in the interstellar medium}, volume={5}, rights={2020 The Author(s), under exclusive licence to Springer Nature Limited}, journal={Nature Astronomy}, author={Ioppolo, S. and Fedoseev, G. and Chuang, K.-J. and Cuppen, H. M. and Clements, A. R. and Jin, M. and Garrod, R. T. and Qasim, D. and Kofman, V. and van Dishoeck, E. F. and Linnartz, H.}, year={2021}, pages={197},  }

@article{Chuang_Fedoseev_Qasim_Ioppolo_Jger_Henning_Palumbo_Van_Dishoeck_Linnartz_2020,
title={Formation of complex molecules in translucent clouds: acetaldehyde, vinyl alcohol, ketene, and ethanol via “nonenergetic” processing of {C$_2$H$_2$} ice}, volume={635}, journal={Astronomy and Astrophysics}, author={Chuang, K. J. and Fedoseev, G and Qasim, D and Ioppolo, S and Jäger, C and Henning, Th and Palumbo, M E and Van Dishoeck, E F and Linnartz, H}, year={2020} }

@article{Oberg_2016, title={Photochemistry and Astrochemistry: Photochemical Pathways to Interstellar Complex Organic Molecules}, volume={116},  journal={Chem. Rev.}, author={Öberg, Karin I.}, year={2016}, pages={9631} }

@article{Oberg_Garrod_van_Dishoeck_Linnartz_2009, title={Formation rates of complex organics in {UV} irradiated {CH$_3$OH}-rich ices {I}. Experiments}, volume={504}, journal={A\&A}, author={Öberg, K I and Garrod, R T and van Dishoeck, E F and Linnartz, H}, year={2009}, pages={891} }

@article{gudipati_situ_2012,
	title = {{In-{situ} {probing} {of} {radiation}-{induced} {processing} {of} {organics} {in} {astrophysical} {ice} {analogs}-{novel} {laser} {desorption} {laser} {ionization} {time}-{of}-{flight} {mass} {spectroscopic} {studies}}},
	volume = {756},
	
	journal = {Astrophys. J. Lett.},
	author = {Gudipati, Murthy S and Yang, Rui},
	year = {2012},
}

@article{turner_exploiting_2020,
	title = {Exploiting Photoionization Reflectron Time-of-Flight Mass Spectrometry to Explore Molecular Mass Growth Processes to Complex Organic Molecules in Interstellar and Solar System Ice Analogs},
	volume = {53},
	journal = {Acc. Chem. Res.},
	author = {Turner, Andrew M. and Kaiser, Ralf I.},
	year = {2020},
	pages = {2791},
}

@article{jones_application_2013,
	title = {Application of Reflectron  Time-of-Flight Mass Spectroscopy in the Analysis of Astrophysically Relevant Ices Exposed to Ionization Radiation: Methane {(CH$_4$)} and {D$_4$}-Methane ({CD}$_4$) as a Case Study},
	volume = {4},
	shorttitle = {Application of {Reflectron} {Time}-of-{Flight} {Mass} {Spectroscopy} in the {Analysis} of {Astrophysically} {Relevant} {Ices} {Exposed} to {Ionization} {Radiation}},
	  
	journal = {J. Phys. Chem. Lett.},
	author = {Jones, Brant M. and Kaiser, Ralf I.},
	  
	year = {2013},
	pages = {1965},
	
}

@article{Paardekooper_Bossa_Isokoski_Linnartz_2014, title={Laser desorption time-of-flight mass spectrometry of ultraviolet photo-processed ices}, volume={85}, journal={Rev. Sci. Instrum.}, author={Paardekooper, D. M. and Bossa, J. B. and Isokoski, K. and Linnartz, H.}, year={2014}, pages={10}}

@article{Bulak_Paardekooper_Fedoseev_Linnartz_2021, title={Photolysis of acetonitrile in a water-rich ice as a source of complex organic molecules: {CH$_3$CN} and {H$_2$O:CH$_3$CN} ices}, volume={647}, journal={Astron. Astrophys}, author={Bulak, M. and Paardekooper, D. M. and Fedoseev, G. and Linnartz, H.}, year={2021}, pages={A82} }

@article{Prasad_Tarafdar_1983, title={UV radiation field inside dense clouds - Its possible existence and chemical implications}, volume={267}, journal={Astrophys. J.}, author={Prasad, S. S. and Tarafdar, S. P.}, year={1983}, pages={603} }

@article{Ligterink_Paardekooper_Chuang_Both_CruzDiaz_Van_Helden_Linnartz_2015, title={Controlling the emission profile of an {H$_2$} discharge lamp to simulate interstellar radiation fields}, volume={584}, journal={Astron. Astrophys}, author={Ligterink, N. F.W. and Paardekooper, D. M. and Chuang, K. J. and Both, M. L. and Cruz-Diaz, G. A. and Van Helden, J. H. and Linnartz, H.}, year={2015}, pages={A56} }

@article{boogert_observations_2015,
	title = {Observations of the {Icy} {Universe}},
	volume = {53},
	journal = {Ann. Rev. Astron. Astrophys.},
	author = {Boogert, Adwin and Gerakines, Perry and Whittet, Douglas},
	  
	year = {2015},
	pages = {541},
		
}

@article{dartois_methanol_1999,
	title = {Methanol: {The} second most abundant ice species towards the high-mass protostars {RAFGL7009S} and {W} {33A}},
	volume = {342},
	shorttitle = {Methanol},
	journal = {Astronomy and Astrophysics},
	author = {Dartois, Emmanuel and Schutte, W. and Geballe, Thomas and Demyk, Karine and Ehrenfreund, Pascale and D'hendecourt, Louis},
	 
	year = {1999},
	pages = {L32},
}

@article{pontoppidan_detection_2003,
	title = {Detection of abundant solid methanol toward young low mass stars},
	volume = {404},
	
	
	journal = {Astron. Astrophys},
	author = {Pontoppidan, K. M. and Dartois, E. and van Dishoeck, E. F. and Thi, W.-F. and d'Hendecourt, L.},
	  
	year = {2003},
	pages = {L17},

	
}

@article{paardekooper_laser_2016,
	title = {Laser desorption time-of-flight mass spectrometry of vacuum {UV} photo-processed methanol ice},
	volume = {592},

	
	  
	journal = {Astron. Astrophys},
	author = {Paardekooper, D. M. and Bossa, J.-B. and Linnartz, H.},
	  
	year = {2016},
	pages = {A67},
	
}

@article{abou_mrad_methanol_2016,
	title = {Methanol ice {VUV} photoprocessing: {GC}-{MS} analysis of volatile organic compounds},
	volume = {458},
	
	shorttitle = {Methanol ice {VUV} photoprocessing},
	
	  
	journal = {Mon. Not. Roy. Astron. Soc.},
	author = {Abou Mrad, Ninette and Duvernay, Fabrice and Chiavassa, Thierry and Danger, Gr{\'e}goire},
	  
	year = {2016},
	pages = {1234},
	
}

@article{Hudgins_Sandford_Allamandola_Tielens_1993, title={Mid- and far-infrared spectroscopy of ices - Optical constants and integrated absorbances}, journal={The Astrophys. J. Suppl. Ser.},  volume={86},  author={Hudgins, D. M. and Sandford, S. A. and Allamandola, L. J. and Tielens, A. G. G. M.}, year={1993}, pages={713}}

@article{chen_soft_2013,
	title = {{Soft} {x}-{ray} {irradiation} {of} {methanol} {ice}: {formation} {of} {products} {as} {a} {function} {of} {photon} {energy}},
	volume = {778},
	
	shorttitle = {{SOFT} {X}-{RAY} {IRRADIATION} {OF} {METHANOL} {ICE}},
	
	journal = {Astrophys. J.},
	author = {Chen, Y.-J. and Ciaravella, A. and Caro, G. M. Mu{\~n}oz and Cecchi-Pestellini, C. and Jim{\'e}nez-Escobar, A. and Juang, K.-J. and Yih, T.-S.},
	  
	year = {2013},
	
	pages = {162},
	
}

@article{cruz_2016,
	title = {Negligible photodesorption of methanol ice and active photon-induced desorption of its irradiation products},
	volume = {592},
	

	journal = {Astron. Astrophys},
	author = {Cruz-Diaz, G. A. and Mart{\'i}n-Dom{\'e}nech, R. and Mu{\~n}oz Caro, G. M. and Chen, Y.-J.},
	  
	year = {2016},
	pages = {A68},
	
}

@article{freitas_laboratory_2020,
	title = {{Laboratory} {investigation} {of} {X}-{ray} {photolysis} {of} {methanol} {ice} {and} {its} {implication} {on} {astrophysical} {environments}},
	volume = {43},


	journal = {Qu{\'i}mica Nova},
	author = {Freitas, Fabricio M. and Pilling, S{\'e}rgio},
	  
	year = {2020},
	pages = {521},
	
}

@article{maity_infrared_2014,
	title = {Infrared and reflectron time-of-flight mass spectroscopic study on the synthesis of glycolaldehyde in methanol ({CH$_3$OH}) and methanol{\textendash}carbon monoxide ({CH$_3$OH}{\textendash}{CO}) ices exposed to ionization radiation},
	volume = {168},

	
	journal = {Faraday Discuss.},
	author = {Maity, Surajit and Kaiser, Ralf I. and Jones, Brant M.},
	  
	year = {2014},
	
	pages = {485},
	
}

@article{maity_formation_2015,
	title = {Formation of complex organic molecules in methanol and methanol{\textendash}carbon monoxide ices exposed to ionizing radiation {\textendash} a combined {FTIR} and reflectron time-of-flight mass spectrometry study},
	volume = {17},
	
	
	journal = {Phys. Chem. Chem. Phys. },
	author = {Maity, Surajit and Kaiser, Ralf I. and Jones, Brant M.},
	 
	year = {2015},
	
	pages = {3081},
	
}

@article{tenelanda-osorio_effect_2022,
	title = {Effect of the {UV} dose on the formation of complex organic molecules in astrophysical ices: irradiation of methanol ices at 20 {K} and 80 {K}},
	volume = {515},
	
	shorttitle = {Effect of the {UV} dose on the formation of complex organic molecules in astrophysical ices},

	journal = {Mon. Not. Roy. Ac. Soc.},
	author = {Tenelanda-Osorio, Laura I and Bouquet, Alexis and Javelle, Thomas and Mousis, Olivier and Duvernay, Fabrice and Danger, Gr{\'e}goire},
	  
	year = {2022},
	pages = {5009},
	
}

@article{henderson_direct_2015,
	title = {{Direct} {detection} {of} {complex} {organic} {products} {in} {ultraviolet} ({Ly$\alpha$}) {and} {electron}-{irradiated} {astrophysical} {and} {cometary} {ice} {analogs} {using} {two}-{step} {laser} {ablation} {and} {ionization} {mass} {spectrometry}},
	volume = {800},


	journal = {Astrophys. J.},
	author = {Henderson, Bryana L. and Gudipati, Murthy S.},
	 
	year = {2015},
	
	pages = {66},
	
}

@article{yang_novel_2014,
	title = {Novel two-step laser ablation and ionization mass spectrometry ({2S}-{LAIMS}) of actor-spectator ice layers: {Probing} chemical composition of {D$_2$O} ice beneath a {H$_2$O} ice layer},
	volume = {140},
	
	shorttitle = {Novel two-step laser ablation and ionization mass spectrometry ({2S}-{LAIMS}) of actor-spectator ice layers},
	
	journal = {J. Chem. Phys.},
	author = {Yang, Rui and Gudipati, Murthy S.},
	 
	year = {2014},
	pages = {104202},
	
}

@article{mathew_methanol_2022,
	title = {Methanol in the {RNA} world: {An} astrochemical perspective},
	volume = {9},

	shorttitle = {Methanol in the {RNA} world},
	
	journal = {Front. Astron. Space Sci.},
	author = {Mathew, Thomas and Esteves, Pierre Moth{\'e} and Prakash, G. K. Surya},
	year = {2022},
	
}

@article{VanGelder_Tabone_Tychoniec_Van_DIshoeck_Beuther_Boogert_Caratti_Garatti_Klaassen_Linnartz_Muller_etal_2020, title={Complex organic molecules in low-mass protostars on Solar System scales - I. Oxygen-bearing species}, volume={639}, journal={Astronomy \&Astrophysics}, author={Van Gelder, M. L. and Tabone, B. and Tychoniec, L. and Van Dishoeck, E. F. and Beuther, H. and Boogert, A. C.A. and Caratti O Garatti, A. and Klaassen, P. D. and Linnartz, H. and Müller, H. S.P. and Taquet, V.}, year={2020}, pages={A87} }

@article{Nazari_VanGelder_Van_Dishoeck_Tabone_VanT_Hoff_Ligterink_Beuther_Boogert_Caratti_Garatti_Klaassen_etal_2021, title={Complex organic molecules in low-mass protostars on Solar System scales - {II}. Nitrogen-bearing species}, volume={650}, journal={Astron. Astrophys}, author={Nazari, P. and Van Gelder, M. L. and Van Dishoeck, E. F. and Tabone, B. and Van ’t Hoff, M. L. R. and Ligterink, N. F. W. and Beuther, H. and Boogert, A. C. A. and Caratti O Garatti, A. and Klaassen, P. D. and Linnartz, H. and Taquet, V. and Tychoniec}, year={2021}, pages={A150} 
}

@article{Munoz_2002,
 title={Amino acids from ultraviolet irradiation of interstellar ice analogues}, volume={416}, rights={2002 Springer Nature Limited}, journal={Nature}, author={Munoz Caro, G. M. and Meierhenrich, U. J. and Schutte, W. A. and Barbier, B. and Arcones Segovia, A. and Rosenbauer, H. and Thiemann, W. H.P. and Brack, A. and Greenberg, J. M.}, year={2002}, pages={403}}

@article{Greenberg_Gillette_Caro_Mahajan_Zare_Li_Schutte_Groot_Mendoza_Gomez_2000, title={Ultraviolet Photoprocessing of Interstellar Dust Mantles as a Source of Polycyclic Aromatic Hydrocarbons and Other Conjugated Molecules}, volume={531}, journal={Astrophys. J. Lett.}, author={Greenberg, J. Mayo and Gillette, J. Seb and Caro, Guillermo M. Muñoz and Mahajan, Tania B. and Zare, Richard N. and Li, Aigen and Schutte, Willem A. and Groot, Menno de and Mendoza-Gómez, Celia}, year={2000}, pages={L71}}

@article{Jin_Garrod_2020, title={Formation of Complex Organic Molecules in Cold Interstellar Environments through Nondiffusive Grain surface and Ice mantle Chemistry}, journal={The Astrophys. J. Suppl. Ser.}, volume={249}, author={Jin, Mihwa and Garrod, Robin T.}, year={2020}, pages={26}}

@article{Luna_Molpeceres_Ortigoso_Satorre_Domingo_Mate_2018, title={Densities, infrared band strengths, and optical constants of solid methanol}, volume={617}, rights={© ESO 2018}, journal={Astron. Astrophys}, author={Luna, Ramón and Molpeceres, Germán and Ortigoso, Juan and Satorre, Miguel Angel and Domingo, Manuel and Maté, Belén}, year={2018}, pages={A116} }

@article{Grace_Butcher_Monroe_Nikkel_2017, title={Index of refraction, {Rayleigh} scattering length, and {Sellmeier} coefficients in solid and liquid argon and xenon}, volume={867}, journal={Nucl. Instrum. Methods Phys. Res. A}, author={Grace, Emily and Butcher, Alistair and Monroe, Jocelyn and Nikkel, James A.}, year={2017}, pages={204} }

@article{Krasnokutski_2022, title={A pathway to peptides in space through the condensation of atomic carbon}, volume={6}, rights={2022 The Author(s)}, journal={Nature Astronomy}, author={Krasnokutski, S. A. and Chuang, K.-J. and Jäger, C. and Ueberschaar, N. and Henning, Th}, year={2022}, pages={381} }

@article{Holtom_2005, title={A Combined Experimental and Theoretical Study on the Formation of the Amino Acid Glycine {(NH$_2$CH$_2$COOH)} and Its Isomer {(CH$_3$NHCOOH)} in Extraterrestrial Ices}, volume={626}, journal={Astrophys. J.}, author={Holtom, Philip D. and Bennett, Chris J. and Osamura, Yoshihiro and Mason, Nigel J. and Kaiser, Ralf I.}, year={2005}, pages={940}}

@article{henderson_plume_2014,
	title = {Plume Composition and Evolution in Multicomponent Ices Using Resonant Two-Step Laser Ablation and Ionization Mass Spectrometry},
	volume = {118},
		
	journal = {J. Phys. Chem. A},
	author = {Henderson, Bryana L. and Gudipati, Murthy S.},
	  
	year = {2014},
	pages = {5454},
	
}

@article{Mason_2014, title={Electron induced chemistry: a new frontier in astrochemistry}, volume={168}, journal={Faraday Discuss.}, author={Mason, Nigel J. and Nair, Binukumar and Jheeta, Sohan and Szymańska, Ewelina}, year={2014}, pages={235}}

@article{Andrade_2009, title={Frozen methanol bombarded by energetic particles: Relevance to solid state astrochemistry}, volume={603}, journal={Surface Science}, author={Andrade, D. P. P. and Boechat-Roberty, H. M. and Martinez, R. and Homem, M. G. P. and da Silveira, E. F. and Rocco, M. L. M.}, year={2009}, pages={1190} }

@article{Almeida_2012, title={Desorption from Methanol and Ethanol Ices by High Energy Electrons: Relevance to Astrochemical Models}, volume={116}, journal={J. Phys. Chem. C}, author={Almeida, Guilherme C. and Andrade, Diana P. P. and Arantes, C. and Nazareth, Andressa M. and Boechat-Roberty, Heloisa M. and Rocco, Maria Luiza M.}, year={2012}, pages={25388} }

@article{de_Marcellus_2015, title={Aldehydes and sugars from evolved pre-cometary ice analogs: Importance of ices in astrochemical and prebiotic evolution}, volume={112}, journal={Proc. Nat. Ac. Sci.}, author={de Marcellus, Pierre and Meinert, Cornelia and Myrgorodska, Iuliia and Nahon, Laurent and Buhse, Thomas and d’Hendecourt, Louis Le Sergeant and Meierhenrich, Uwe J.}, year={2015}, pages={965} }

@article{Dartois_2020, title={Non-thermal desorption of complex organic molecules: Cosmic-ray sputtering of {CH$_3$OH} embedded in {CO$_2$} ice}, volume={634}, journal={Astron. Astrophys}, author={Dartois, E. and Chabot, M. and Bacmann, A. and Boduch, P. and Domaracka, A. and Rothard, H.}, year={2020}, pages={A103}}

@article{Munoz_Caro_Ciaravella_2019, title={X-ray versus Ultraviolet Irradiation of Astrophysical Ice Analogs Leading to Formation of Complex Organic Molecules}, volume={3}, journal={ACS Earth and Space Chemistry}, author={Muñoz Caro, Guillermo M. and Ciaravella, Angela and Jiménez-Escobar, Antonio and Cecchi-Pestellini, Cesare and González-Díaz, Cristóbal and Chen, Yu-Jung}, year={2019},  pages={2138} }

@article{Gibb_Whittet_Boogert_Tielens_2004, title={Interstellar Ice: The Infrared Space Observatory Legacy}, volume={151}, journal={Astrophys. J. Suppl. Ser.}, author={Gibb, E. L. and Whittet, D. C. B. and Boogert, A. C. A. and Tielens, A. G. G. M.}, year={2004}, pages={35}}

@article{Bernstein_1995, title={Organic Compounds Produced by Photolysis of Realistic Interstellar and Cometary Ice Analogs Containing Methanol}, volume={454}, journal={Astrophys. J.}, author={Bernstein, Max P. and Sandford, Scott A. and Allamandola, Louis J. and Chang, Sherwood and Scharberg, Maureen A.}, year={1995}, pages={327} }

@article{Focsa_Mihesan_Ziskind_Chazallon_Therssen_Desgroux_Destombes_2006, title={Wavelength-selective vibrationally excited photodesorption with tunable IR sources}, volume={18}, journal={Journal of Physics: Condensed Matter}, author={Focsa, C. and Mihesan, C. and Ziskind, M. and Chazallon, B. and Therssen, E. and Desgroux, P. and Destombes, J. L.}, year={2006}, pages={S1357}}

@article{Bulak_Paardekooper_Fedoseev_2021, title={Photolysis of acetonitrile in a water-rich ice as a source of complex organic molecules: {CH$_3$CN} and {H$_2$O:CH$_3$CN} ices}, volume={647}, journal={Astron. Astrophys}, author={Bulak, M. and Paardekooper, D. M. and Fedoseev, G. and Linnartz, H.}, year={2021}, pages={A82} }

\end{document}